\documentstyle[pra,aps,epsfig,amsmath,amssymb]{revtex}
\begin{document}
\draft
\title
{\bf Bose-Einstein condensates at equilibrium inside a pancake-shaped
trap: dimensional cross-over in the scattering properties}
\author{B. Tanatar$^{(1)}$, A. Minguzzi$^{(2)}$, P. Vignolo$^{(2)}$
and M.P. Tosi$^{(2)}$}
\address{
$^{(1)}$Department of Physics, Bilkent University, Bilkent, 06533
Ankara, Turkey\\
$^{(2)}$INFM and Classe di Scienze, Scuola Normale Superiore, I-56126
Pisa, Italy}
\maketitle
\begin{abstract}
Two-dimensionality of the scattering events in a Bose-Einstein condensate
introduces a logarithmic dependence on density in the coupling constant
entering a mean-field theory of the equilibrium density profile, 
which becomes dominant
as the $s$-wave scattering length gets larger than the condensate thickness.
We analyze quantitatively the role of the form of the coupling constant in 
determining the transverse profile of a condensate confined in a
harmonic pancake-shaped trap at
zero temperature. We trace the regions of experimentally
accessible system parameters for which the cross-over between 
different dimensionality
behaviors may become observable through {\it in situ} imaging of the condensed
cloud with varying trap anisotropy and scattering length.
\end{abstract}



\section{Introduction}

A system of dilute hard-disk bosons constrained to move inside a
two-dimensional (2D)  box is a well-known model in statistical mechanics
\cite{schick}. This system shows very different features from 
 its three-dimensional (3D) counterpart. First of all its collisional
properties are very peculiar, since in two dimensions the
T-matrix vanishes at low momentum and energy
\cite{schick,popov}, and the resulting boson-boson coupling constant
depends on the density of the system through the logarithm of the
diluteness parameter $n a^2$, where $n$ is the areal density of the disks
and $a$ the
hard-core diameter. Another important difference with respect to 3D systems is
that while at zero temperature a fraction of the bosons undergo
Bose-Einstein condensation, at finite temperature phase fluctuations
destroy the long-range order (in agreement with the Mermin-Wagner
theorem \cite{mermin-wagner}) and the one-body density matrix decays
algebraically to zero at large distances. The interacting
2D Bose gas is nevertheless a superfluid also at finite
temperature and the nature of the transition is known to be of the 
Kosterlitz-Thouless type \cite{KT}, the critical temperature being
$T_{KT}\approx \hbar^2 n/(m\ln \ln (1/na^2))$ in the ultra-dilute
limit $\ln \ln (1/na^2)\gg 1$ \cite{HF}. In a recent experiment a
2D Bose gas has been realized using a film of gaseous Hydrogen on a Helium
substrate \cite{jakkola}.

A dilute 2D Bose gas inside a harmonic trap has also been
considered by several authors. For the ideal gas the
presence of the confinement introduces essential changes
in the thermodynamic
properties relative to the homogeneous case, and Bose-Einstein
condensation starts at a finite temperature $k_BT_c\approx \hbar
\omega_\perp N^{1/2}$ where $\omega_\perp$ is the frequency of the 2D
harmonic confinement and $N$ the number of particles\cite{ideal_gas}.
The interacting system is expected to be
Bose-condensed at low temperature $T<T_\phi$, but phase fluctuations
should become important at a temperature $T_\phi$ where
the phase correlation length becomes smaller than the radius of the
cloud \cite{petrov}. The precise behavior of the system at these
temperatures and the nature of the phase transition have not yet been
fully clarified (see {\it e.g.} Fernandez and Mullin \cite{mullin} 
and references
therein). 

In current experiments on Bose-Einstein condensates of alkali
atoms in magnetic or optical traps the anisotropy of the confinement
can be varied to obtain flatter and flatter (quasi-2D) 
condensates~\cite{bec2d}, with the ultimate aim to observe
the special features of low dimensionality. It is thus important
to realize where in parameters space and which physical aspects
of 2D systems will become dominant as the anisotropy
is increased. The key question in this respect is which
is the appropriate model for the boson-boson coupling strength in the
regime of cross-over between 3D and 2D
 and how to insert it in
a self-consistent calculation of the equilibrium properties 
of a Bose-Einstein condensed cloud. 

In this paper we treat this problem at zero temperature. We
evaluate the equilibrium density profiles of the condensate under
different choices for its physical parameters, ranging from a
3D anisotropic system to a strictly 2D
one. With increasing anisotropy the system first becomes 
2D with regard to the confinement -- only the lowest
axial state is occupied -- and then also in its
collisional properties. We find that these
different regimes can be identified by observing the size of the
cloud in the radial plane, and we characterize the cross-over in terms
of the relevant physical parameters.

The paper is organized as follows. In Sec.~\ref{model}  we introduce
the coupling strengths used to describe the condensate from the 3D to the
2D regime and the corresponding non-linear Schr\"odinger equation (NLSE).
The results for the equilibrium density profile
are given in Sec.~\ref{results}, and
Sec.~\ref{conclusions} presents our conclusions and final remarks.

\section{Ground state and scattering properties of a quasi-2D condensate}
\label{model}

We consider a dilute Bose-condensed gas at zero temperature under anisotropic
(pancake-shaped) harmonic confinement characterized by the frequencies
$\omega_{\perp}$ and $\omega_z=\lambda\omega_{\perp}$ with $\lambda\gg 1$.
Within a density-functional approach, the ground-state properties of
the gas are described in terms of the condensate wave 
function $\psi(r)$ in the $\{x,y\}$ plane, which is to be determined as the 
minimum of the local-density energy functional
\begin{equation}
E[\psi]=\int d^2r\left[\frac{\hbar^2}{2m}|\nabla\psi|^2+(V_{ext}-
\mu)|\psi|^2+\epsilon(n)|\psi|^2\right].
\end{equation}
Here $n=|\psi|^2$ is the particle density, $\mu$ is the chemical potential,
and $\epsilon(n)=gn/2$ is the ground-state energy per particle of a homogeneous
Bose gas with short-range interactions in the mean-field approximation.
As we shall see, the coupling $g$ can depend on the condensate density: 
$g\equiv g(\psi)$.
The Euler equation $\delta E/\delta\psi^*=0$ leads to the NLSE
for the condensate wave function\cite{nota1}
\begin{equation}
-\frac{\hbar^2}{2m}\nabla^2\psi+V_{ext}\psi+g(\psi)|\psi|^2\psi=\mu\psi.
\label{nlse}
\end{equation}
The familiar Gross-Pitaevskii equation is recovered by taking a constant
coupling $g$.
Extensions of this equation beyond mean field and beyond the local-density
approximation have been proposed for systems at higher density
\cite{Fabrocini,Nunes,Andersen}.

We now need to specify the choice of $g$ for the system of present interest.
Quite generally the coupling $g$ is obtained microscopically from the
effective interaction potential $\Gamma(k,k',P)$ in the limit of low energy
and momenta. The effective interaction can in turn be related to the 
two-body scattering function $f(\vec{k},\vec{k'})=\int\psi_{\vec{k}}(r)v(r)
e^{-i\vec{k'}\cdot\vec{r}} d^dr$ in dimension $d=3$ or 2, 
with $\psi_{\vec{k}}(r)$
 being the outgoing wave function for the relative motion of the two
particles and $v(r)$ being the interparticle potential \cite{Abrikosov}.
For a 3D system $f(\vec{k},\vec{k'})$ is the scattering 
amplitude and $g$ is a constant determined by the $s$-wave scattering 
length $a$:
\begin{equation}
g_{3D}=\frac{4\pi\hbar^2}{m}a.
\end{equation}
On increasing the anisotropy of a 3D condensate, its physical
properties will change first due to the modified shape of the confinement
\cite{Salasnich}, but then also due to the modified scattering properties.
We consider a condensate in a pancake-shaped trap which is flat enough
so that the dimension of the cloud in the axial direction is of the order
of the harmonic oscillator length $a_z=\sqrt{\hbar/
m\omega_{z}}$.
If the condition $a\ll a_z$ holds, the system experiences collisions in
three dimensions and the coupling constant to be used 
in the 2D NLSE is 
\begin{equation}
g_{Q3D}=g_{3D}|\phi_0(0)|^2
\label{q3d}
\end{equation}
where $\phi_0(0)=(2\pi a_z^2)^{-1/4}$ is the axial ground-state wave function
evaluated at $z=0$.
This coupling constant has been used by a number of authors (see {\it e.g.} 
\cite{15} and references therein).
When the anisotropy further increases and $a$ approaches $a_z$, 
the collisions start
to be influenced by the presence of the trap in the tight $z$ direction
and the coupling $g$ in such a quasi-2D condensate is to be taken in the
form
\begin{equation}
g_{Q2D}=g_{3D}|\phi_0(0)|^2\frac{1}{1+\dfrac{a}{a_z\sqrt{2\pi}}
|\ln{(2(2\pi)^{3/2}naa_z)}|}.
\label{q2d}
\end{equation}
This expression was originally derived by Petrov {\it et
al.}~\cite{petrov,petrov2} by studying the scattering 
function of a system which is harmonically confined in the $z$ direction
and homogeneous in the $\{x,y\}$ plane (see also footnote [18] in
\cite{petrov}). 
The coupling $g$ in this case depends on  density, as is typical of 
2D collisions.

Indeed, when the collisions are in the fully 2D regime ($a\gg a_z$), the
system is described by the coupling
\begin{equation}
g_{2D}=\frac{4\pi\hbar^2}{m}\frac{1}{|\ln{(na^2)}|},
\label{2d}
\end{equation}
as was first derived by Schick for the homogeneous 2D system \cite{schick}.
The use of the coupling $g_{2D}$  for a system under external
confinement, involving a dependence on the local
density,  has been proposed by Shevchenko~\cite{shevchenko} and
more recently 
by Kolomeisky {\it et al.}~\cite{14}. 
It was   rigorously justified by Lieb and coworkers~\cite{15}.

In the following Section we will present results for the condensate wave
function $\psi(r)$ as obtained from the NLSE (\ref{nlse}) using the three
couplings $g_{Q3D}$, $g_{Q2D}$ and $g_{2D}$.

\section{Equilibrium density profiles}
\label{results}

In  current experiments on Bose-Einstein condensates of alkali
atoms, one of the observables which are most directly accessible is the 
density profile.
In particular, it is possible to take {\it in situ} images of the cloud, which can 
be directly compared with the results of theoretical models.
We give in this Section the predictions for the widths and the shapes of 
the equilibrium density in the $\{x,y\}$ plane for a 
condensate in a pancake-shaped trap under different collisional regimes 
as the anisotropy is increased.

Before proceeding to the numerical solution of the NLSE (\ref{nlse}),
 it is useful to give  a simple analytical estimate
of the expected width of the cloud. We use for this purpose the Thomas-Fermi approximation,
{\it i.e.} we neglect the kinetic-energy term in Eq. (\ref{nlse}) to obtain the
density profile
\begin{equation}
|\psi_{TF}(r)|^2=\frac{1}{\tilde{g}(\mu)}(\mu-V_{ext}(r)) 
\theta(\mu-V_{ext}(r)).  
\label{TFA}
\end{equation}
Here we have also neglected the spatial dependence of the coupling constant
$\tilde{g}$ and used the result of the homogeneous system to relate the 
density to the chemical potential \cite{nota3}.
Equation (\ref{TFA}) is valid when the number of atoms in the condensate 
is large and in this regime we have found that its predictions agree 
well with the full numerical solution of Eq. (\ref{nlse}).

The chemical potential in Eq. (\ref{TFA}) is fixed by imposing the 
normalization condition $N=\int|\psi_{TF}(r)|^2 d^2r$. By inversion
of this equation we find an expression for the Thomas-Fermi width
of the cloud, $R_{TF}=\sqrt{2\mu/m\omega_{\perp}^2}$, in terms of
the physical parameters of the system. We show in Fig.~\ref{figTF}
the width of the cloud as a function of the ratio $a/a_z$
for a given choice of the number of atoms and of the anisotropy
($N=5\times 10^5$,  $\lambda=2\times 10^5$).
We expect the quasi-3D model (\ref{q3d}) for the coupling to be 
accurate only for small values for $a/a_z$. It evidently overestimates the size
of the cloud as $a/a_z$ increases.
The quasi-2D behavior should be correct for $a/a_z>0.1$ and is ultimately
superseded by the purely 2D behavior as $a/a_z$ becomes appreciably 
larger than unity.

We now turn to the numerical solution of Eq.~(\ref{nlse}) with the
alternative expressions (\ref{q3d})-(\ref{2d}) for the coupling.
We have used the steepest descent method\cite{33,Fabrocini}, which is
known to produce accurate results. A further check of our
numerical calculations is provided by the virial relation.

We give illustrations of the ground-state wave function $\psi(r)$
predicted by the three models in some relevant cases.
First of all we have considered the values of particle number, anisotropy
parameter, and scattering length as appropriate for $^{23}$Na atoms in
the experiment of G{\"o}rlitz {\it et al}.\cite{bec2d} ($N=5\times 10^5$, 
$\lambda=26.33$, $a=2.8$ nm). In this case the system is approaching 
two-dimensionality
for what concerns the confinement ($\mu\simeq 2.08 \hbar\omega_z$),
but still is 3D for  collisions ($a/a_z\simeq 3.8\times 10^{-3}$).
As is shown in Fig.~\ref{fig1_BT}, the quasi-3D and  quasi-2D models give
almost indistinguishable predictions (solid and dotted lines). The fully 2D
model would give a quantitatively very different profile (dashed line in Fig. 
\ref{fig1_BT}) with a much larger chemical potential
($\mu=17.8\hbar\omega_z$), but is evidently inapplicable in this regime 
of parameters.

For a second case we increase the anisotropy parameter to 
$\lambda=2\times 10^5$  and make the choice $a/a_z= 0.33$
for the scattering length. This corresponds to the point
of cross-over in the scattering properties from 3D to 2D in a condensate
where motion in the third direction is completely frozen by the confinement
($\mu\simeq 0.002\,\hbar\omega_z$).
In this case (Fig. \ref{fig2_BT}) the three models predict comparable 
shapes of the cloud, which is an indication that indeed the condensate is
entering 
the regime of 2D collisions.

Finally, we have further increased the value of the scattering length
to $a/a_z\simeq 2.68$, keeping the same values for $\lambda$ and
$N$. For this choice of parameters the collisions are truly 2D and the
fully 2D model should give the most accurate prediction for the shape
of the cloud, while the other models overstimate its width
(Fig.~\ref{fig3_BT}).   The calculated chemical potential of the
$g_{2D}$ model is $\mu\simeq 0.0031\,\hbar\omega_z$.

\section{Conclusions}
\label{conclusions}

In summary, we have considered  Bose-Einstein condensates 
confined inside pancake-shaped traps 
at zero
temperature  
within a mean-field description.
Depending on the anisotropy of the trap, we have identified three
different physical regimes for the scattering
events: (i) a quasi-3D regime where the axial dimension of the
condensate is much larger than the scattering length and the
collisions are as in a 3D condensate; (ii) a quasi-2D regime 
where the tight harmonic confinement in the $z$ 
direction begins to influence the scattering events; and (iii)
a fully 2D regime where collisions are restricted to the $\{x,y\}$ plane. 
The 
atom-atom coupling is different in the three cases and a
logarithmic dependence on the density arises
as the 2D effects start to affect the
scattering events. We have adopted a local-density approximation
to introduce the appropriate atom-atom coupling into the energy
functional of the condensate and to obtain a non-linear Schr\"odinger
equation for its in-plane wave function. We have given
numerical solutions of the NLSE for various choices of parameters
corresponding to the three  collisional regimes.  We have
shown that the different collisional regimes are reflected in the 
width of the cloud, which can also be predicted with reasonable 
accuracy by the Thomas-Fermi model if the number of atoms 
in the condensate is not too low. Our results for the fully 2D case
are in accord with those reported recently by Lee {\it et al.}
\cite{burnett} by a numerical solution of the NLSE with the choice
$\tilde g_{2D}(\mu)$ for the coupling strength. 

In conclusion,  
the several interesting properties that are expected  for flat
condensates should be 
observable when not only the condition $\mu<\hbar\omega_z$ is
satisfied (the condensate is 2D for what concerns the confinement), 
but the condition $a/a_z>1$ also holds (the condensate is 2D for what concerns 
the collisions).
The strictly 2D limit can be reached in a pancake-shaped Bose-Einstein
condensate by making the perpendicular
confinement very tight and/or by increasing
the scattering length, {\it e.g.} via Feshbach resonances.
In order to understand the significance of interactions and
high-density effects, high-precision Monte Carlo simulations would be
useful as a test of the range of validity of the mean-field 
and local-density approximations.
Our analysis should be the starting point for
finite-temperature calculations, in which to study the role of the
interactions with the non-condensed thermal particles and the
transitions in phase correlations that are expected to occur with
increasing temperature.

\acknowledgments{
This work was partially supported by INFM under the project
PRA2001-Photonmatter and under the PAIS2000 ``Theory of 
two-dimensional interacting Bose gas''.
B.T. acknowledges support 
from the Scientific and Technical Research 
Council of Turkey (TUBITAK), NATO, the Turkish Department of 
Defense, and the Turkish Academy of Sciences (TUBA), and thanks
Scuola Normale Superiore for hospitality during part of this work. 
A.M. acknowledges
a travel grant from INFM under the initiative ``Calcolo Parallelo''.

\begin{figure}
\centerline{\epsfig{figure=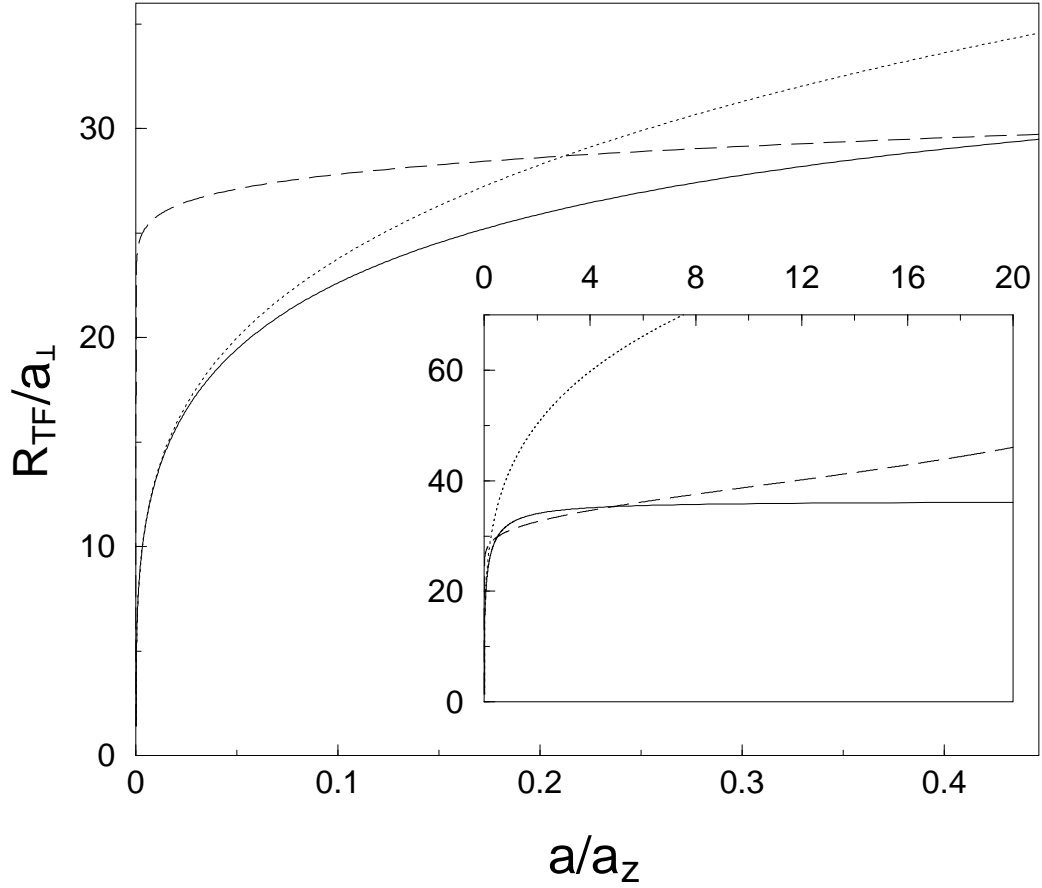,width=1.\linewidth,angle=270}}
\caption{Transverse width of the condensate (in units of
$a_{\perp}=\sqrt{\hbar/m\omega_{\perp}}$) as a function of $a/a_z$
as evaluated in the Thomas-Fermi 
approximation for the three models Q3D (dotted line), Q2D (solid line)
and 2D (dashed line). In the inset the same curves are plotted over a
wide range of the ratio $a/a_z$. The system parameters are
$N=5\times 10^5$ atoms and $\lambda=2\times 10^5$ for the trap
 anisotropy. 
}
\label{figTF}
\end{figure}

\begin{figure}
\centerline{\epsfig{file=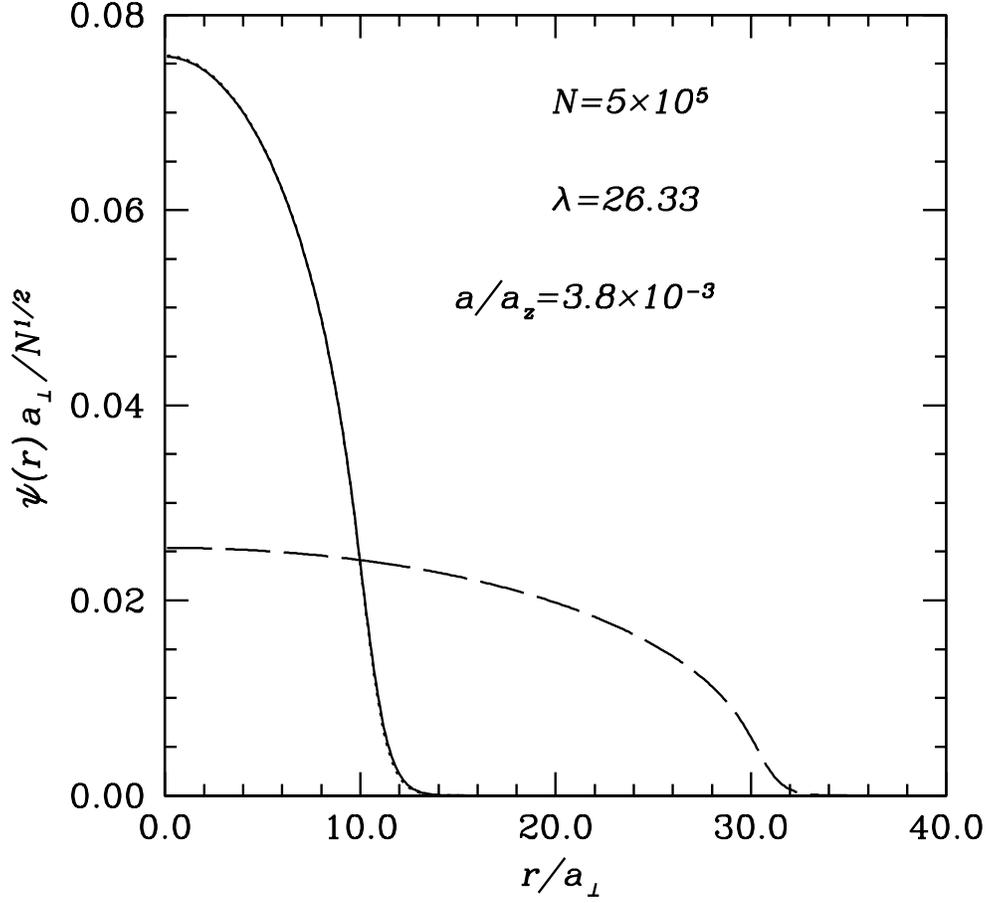,height=1.\linewidth}}
\caption{Condensate wave function $\psi(r)$ (in units of
$\sqrt{N}/a_\perp$) as a function of $r/a_{\perp}$,  for
$a/a_z= 3.8\times 10^{-3}$ 
from the full numerical solution of Eq. (\ref{nlse}) for
the three models Q3D (dotted line), Q2D (solid line)
and 2D (dashed line). 
The values of the particle number and of the anisotropy
parameter are indicated in the figure.}
\label{fig1_BT}
\end{figure}

\begin{figure}
\centerline{\epsfig{figure=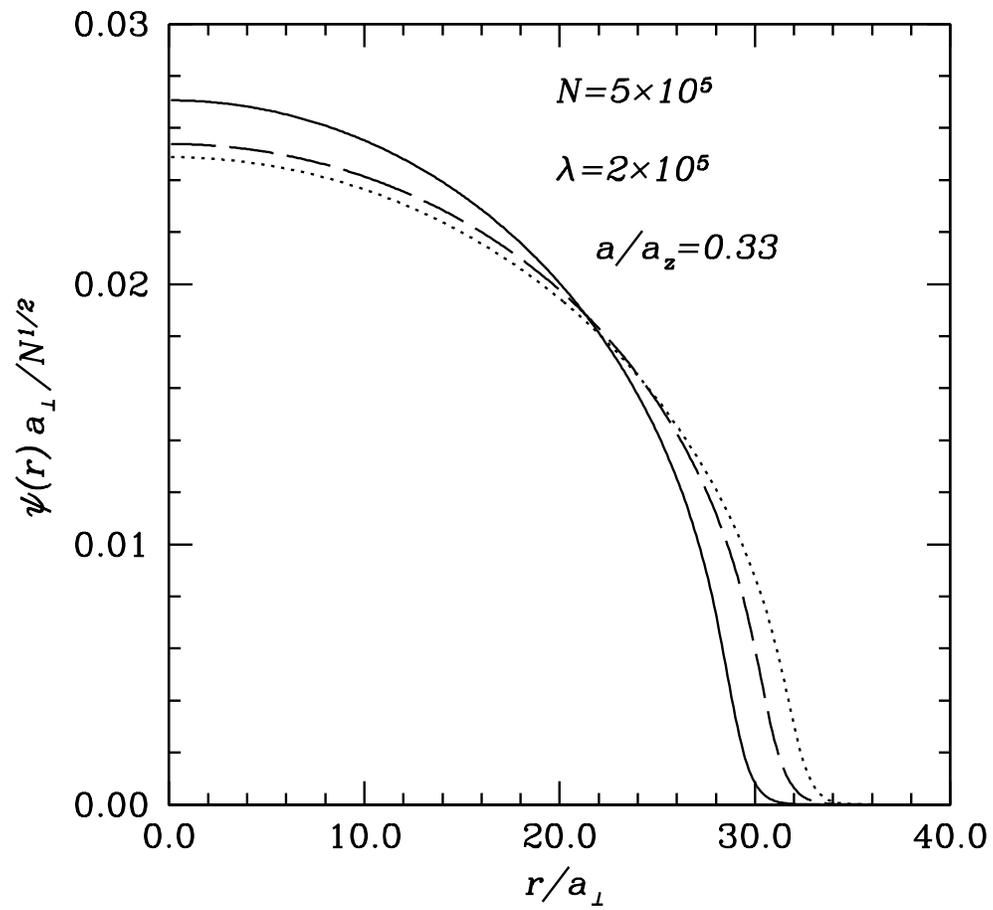,height=1.\linewidth}}
\caption{The same as in Fig.~\ref{fig1_BT}, for $a/a_z= 0.33$
and $\lambda=2\times 10^5$.
}
\label{fig2_BT}
\end{figure}

\begin{figure}
\centerline{\epsfig{figure=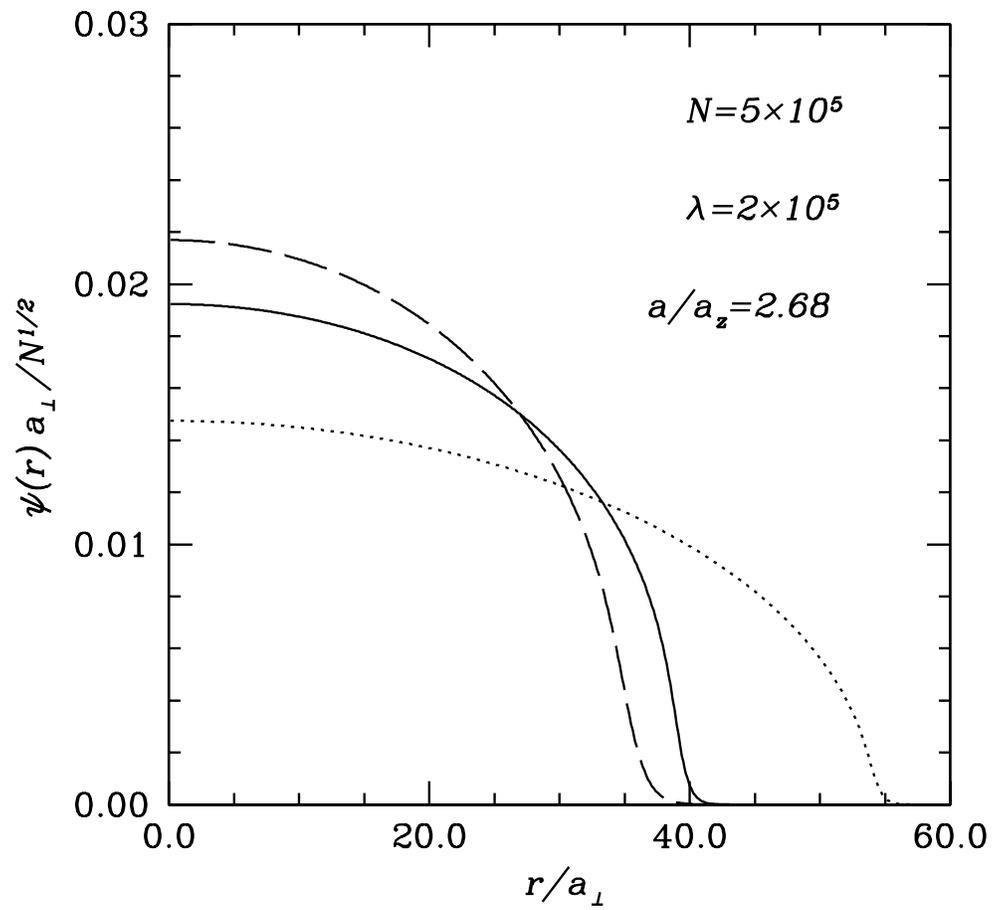,height=1.\linewidth}}
\caption{The same as in Fig.~\ref{fig1_BT}, for $a/a_z= 2.68$
and $\lambda=2\times 10^5$.
}
\label{fig3_BT}
\end{figure}


\begin{references}
\bibitem{nota1}
In the dilute-gas limit we also neglect the higher-order terms 
involving $\delta g/
\delta \psi^*$.
\bibitem{nota3}
The expressions for the coupling $\tilde{g}(\mu)$ that we actually use in
this calculation are
$\tilde{g}_{Q3D}(\mu)=2\sqrt{2\pi}\hbar^2a/(ma_z)$,
$\tilde{g}_{Q2D}(\mu)=(4\pi\hbar^2/m)\,(\sqrt{2\pi}a/a_z+\ln
{(\lambda\hbar\omega_{\perp}/2\pi\mu)})^{-1}$ and 
$\tilde{g}_{2D}(\mu)=(4\pi\hbar^2/m)\,|\ln{(m\mu a^2/4\pi\hbar^2)}|^{-1}$.
\end{references}
\end{document}